\documentclass[twocolumn,noshowpacs]{revtex4}

\usepackage{graphicx}%
\usepackage{dcolumn}
\usepackage{amsmath}
\usepackage{bm}
\usepackage{braket}

\makeatletter
\def\btt#1{\texttt{\@backslashchar#1}}%
\DeclareRobustCommand\bblash{\btt{\@backslashchar}}%
\makeatother

\begin{document}


\title{Quantum correlation between a particle and potential well or barrier}

\author{F.V. Kowalski and R.S. Browne}
\affiliation{Physics Department, Colorado
School of Mines, Golden CO. 80401 U.S.A.}

\begin{abstract}
A two-body quantum correlation is calculated for a particle and an infinite potential well in which it is trapped or either a barrier or finite well over which it traverses. Correlated interference results when the incident and reflected particle substates and their associated well or barrier substates overlap. Measurement of the particle in this region causes a splitting of the well or barrier substate at subsequent times. The joint probability density, which is a function both of the different positions and different times at which the particle and well or barrier are measured, is derived assuming that no interaction occurs between the time each is measured.
\end{abstract}


\maketitle

\section{Introduction}
\label{sec:intro}

Correlation and interference distinguish quantum from classical physics. The former is manifest in the measurement of many-body coincidences predicted by a quantum joint probability density function (PDF). The latter is most familiar as a one-body PDF for an outcome that can be achieved in at least two indistinguishable ways. However, interference can also be generated by superposing many-body states in indistinguishable ways \cite{Gottfried}. 

Quantum correlation between a particle and the center of mass (cm) of the potential well in which it is trapped or the cm of a barrier over which it traverses is described here. When the incident and reflected two-body wavefunctions overlap, correlated interference occurs. This is calculated for the particle and well or barrier all having non-zero rest mass and with motion in free space along one dimension. 

A single-body treatment of a particle in both of these potentials is familiar from a first exposure to quantum mechanics. However, the two-body quantum effects on the particle and cm of the well or barrier, including correlated interference and the ``kinematic'' effects (e.g. recoil) due to the interaction, have not been discussed in the literature as far as we know. We also develop a formalism to describe asynchronous measurement of the particle and well or barrier, the predictions of which are not needed in a single-body treatment.

For a particle traversing a barrier, in a single-body treatment, a refractive index is often used to parametrize the wavefunction; an example of which is the neutron wavefunction traversing a barrier consisting of a slab of matter \cite{rauch}. The analogous classical example is given by an electromagnetic wave traversing a slab of glass, where again an index of refraction is used to parametrize the interaction. The controversy involved with the form of the electromagnetic stress-energy tensor for the electromagnetic wave interacting with matter has a long history and involved many well known physicists of the twentieth century \cite{gordon,loudon}. The analogous quantum result, presented here for the two-body solution, is shown to involve a simple division of kinetic energy and momentum between the particle and barrier.   

A simpler yet similar two-body problem occurs for reflection of a particle from a mirror \cite{kowalski}. Correlated interference in such reflection is an example of perhaps the simplest interferometer, utilizing neither division of amplitude nor division of wavefront methods to generate interference, with path lengths which need not be carefully matched for interference to be manifest. The two-body interaction of a particle with a barrier has effects similar to the particle-mirror interaction: transfer of coherence and the possibility of extending the quantum-classical boundary to a macroscopic mass. Here, however, the focus is on correlated interference and the predictions of synchronous and asynchronous measurements due to the interaction of the particle with the well or barrier. The consequences for asynchronous measurements of a particle reflecting from a mirror have been discussed \cite{kowalski2}. 

A synopsis of this work is as follows. First an overview is given to place the model for calculating asynchronous correlation in context. This model is then shown to conserve probability. Next energy eigenstate solutions to the Schr\"odinger equation for both the well and barrier potentials, modified to include non-simultaneous measurements, are derived. The superposition principle is then used to form reflected particle-well or particle-barrier wavegroups. Predictions from this non-simultaneous joint PDF in the limit of simultaneous measurements are described. This is then related to the predictions for measurement of the particle followed by measurement of the well or barrier. Particularly emphasis is on the correlated interference or overlap region of the incident and reflected particle-well or particle-barrier wavegroups where it is shown that asynchronous measurement of the particle acts as a ``beamsplitter'' for the well or barrier.

\section{Theory}
\label{sec:Theory}

\subsection{Overview}
\label{sec:overview}

A common example of quantum correlation utilizes conservation of angular momentum in particle decay for an isolated system in an eigenstate of angular momentum \cite{Gottfried}. A measurement of the angular momentum of one decay product then always results in the second exhibiting an angular momentum consistent with the initial angular momentum state of the system. Since this property of the second particle can be measured simultaneously with that of the first it is often used as fodder for a discussion about transmitting signals faster than the speed of light \cite{yin}.

By analogy, one might wonder if this ``instantaneous collapse'' extends to non-simultaneous measurements. If so, predictions would then be determined by the following sequence: measurement of one particle is followed by the immediate collapse of the other particle's substate, this second particle's substate then time evolves from its collapsed state, and finally another measurement is made on only the second particle. Such reasoning is based on the assumption that measurement of the first particle collapses the substate of the second particle. We argue that this assumption is incorrect: a measurement of one substate does not collapse all substates of the system.

A more concrete illustration utilizes the angular momentum example mentioned above. Let the initial {\em system} not be in an eigenstate of angular momentum. A measurement on one particle then does not require a unique angular momentum for the second particle since the system is not in an eigenstate of angular momentum. Yet by definition, collapse produces only one of the eigenstates of the operator corresponding to the measured observable. If an angular momentum measurement on the first particle's substate yields a particular value then into what unique value of angular momentum does the second particle collapse?  There never was a well defined angular momentum of the {\em system} to constrain the answer. Yet the answer is needed, in the procedure described above, as an initial condition to determine the time evolution of the second particle until it is actually measured. However, only a measurement of the angular momentum of both particles collapses the wavefunction of the system into a unique angular momentum state and therefore satisfies the definition of a measurement of the system. The sum of simultaneous measurements on each particle yields the angular momentum of the system.

In forming a model which does not rely on the assumption of simultaneous collapse it is useful to review the procedure for determining the results of a measurement of each position simultaneously in a two-body system. This is given by the two-body joint PDF integrated over the spatial extent of each detector. Building on this result, we propose that a measurement of the position of the two particles at different times is given by the joint PDF evaluated over the spatial extent of each detector at the times each is measured. Note that the system is measured only once. There is no collapse and re-measurement of either particle nor of the system. To prevent correlations from occurring after the first particle is measured it is assumed that there is no interaction during the time evolution between measurements.

To account for different temporal measurements of the particle and well or barrier, respective time parameters $t_{1}$ and $t_{2}$ are introduced. These are used both to distinguish between the energies of the particle and the well or barrier in the expression for the wavefunctions (as opposed to using only the total energy) and to label the different times at which the particle and well or barrier are measured. They do not indicate evolution of the subsystems via different Hamiltonians as do the different time variables used by McGuire \cite{Mcguire}. They neither tick at different rates nor are they out of phase, but rather act as only a label, just as $x_{1}$ and $x_{2}$ label the particle and well or barrier spatial coordinates along the $x$ axis. Using particle and well or barrier coordinates $x_{1}$ and $x_{2}$, masses $m$ and $M$, and initial velocities $v$ and $V$ respectively, the two-body Schr\"odinger equation,
\begin{eqnarray}
(\hbar \partial_{x_{1}}^{2}/2m+\hbar \partial_{x_{2}}^{2}/2M + PE[x_{1}-x_{2}] \notag \\
+i\partial_{t})\Psi[x_{1},x_{2},t]=0
\label{eq:Scheqn1time}
\end{eqnarray}
is then written as
\begin{eqnarray}
(\hbar \partial_{x_{1}}^{2}/2m+\hbar \partial_{x_{2}}^{2}/2M+ PE[x_{1}-x_{2}] \notag \\
+i\partial_{t_{1}}+i\partial_{t_{2}})\Psi[x_{1},x_{2},t_{1},t_{2}]=0,
\label{eq:Scheqn2times}
\end{eqnarray}
where the square brackets are used to indicate the argument of a function. The difference between these equations is that the time rate of change of $\Psi$ in the first equation, proportional to the total energy of the system, is separated into a term proportional to the energy of the particle {\em plus} that of the well or barrier in the second equation. The sum of the two time derivative terms in equation \ref{eq:Scheqn2times} is equal to the one time derivative term in equation \ref{eq:Scheqn1time}. Both equations express conservation of energy but parse it differently.

To illustrate the utility of such notation consider calculating the expectation value of the particle's energy using one-time notation, written as $i \hbar \Braket{\Psi|\partial /\partial t |\Psi}$. This of course yields a total energy of the system rather than that of the particle and therefore is not what would be measured. However, $i \hbar \Braket{\Psi|\partial /\partial t_{1} |\Psi}$ gives the appropriate energy expectation value of the particle. A similar result follows for the well or barrier.

The probability of measuring the particle at $(x_{1},t_{1})$ and the well or barrier at $(x_{2},t_{2})$ is then given by $\iint PDF[x_{1},t_{1},x_{2},t_{2}] dx_{1} dx_{2}$ with the joint PDF determined by the solution of equation \ref{eq:Scheqn2times} as $\Psi \Psi^{*}$. Measurement of the particle fixes $(x_{1},t_{1})$ as the PDF evolves along $(x_{2},t_{2})$. Examples of such asynchronous predictions are shown below in figs. \ref{fig:barrierAsynchPeakB} and \ref{fig:WellStandingWave} for the particle-finite well and particle-infinite well.

Conservation of probability can then be expressed locally as,
\begin{eqnarray}
\partial_{t_{1}}PDF[x_{1},t_{1},x_{2},t_{2}] +\partial_{t_{2}} PDF[x_{1},t_{1},x_{2},t_{2}] +\notag \\
\partial_{x_{1}} j_{1}[x_{1},t_{1},x_{2},t_{2}] +\partial_{x_{2}} j_{2}[x_{1},t_{1},x_{2},t_{2}]=0,
\label{eq:consProb}
\end{eqnarray}
where $j_{1}[x_{1},t_{1},x_{2},t_{2}]=\hbar (\Psi^{*} \partial_{x_{1}} \Psi-\Psi \partial_{x_{1}} \Psi^{*})/(2 i m)$ and  $j_{2}[x_{1},t_{1},x_{2},t_{2}]=\hbar (\Psi^{*} \partial_{x_{2}} \Psi-\Psi \partial_{x_{2}} \Psi^{*})/(2 i M)$. While the expressions for these current densities appear similar to that for one particle systems there are subtle but important differences for a two body system \cite{currentdensity}.

Multiplying equation \ref{eq:consProb} by $dx_{1} dx_{2}$, integrating over the segment from $a$ to $b$ along the x-axis ($a \leq x_{1} \leq b$ and $a \leq x_{2} \leq b$), and then rearranging terms yields a solution to equation \ref{eq:consProb} if
\begin{eqnarray}
\partial_{t_{1}} \int_{a}^{b} PDF[x_{1},t_{1},x_{2},t_{2}] dx_{1} +\notag \\ j_{1}[b,t_{1},x_{2},t_{2}]-j_{1}[a,t_{1},x_{2},t_{2}]=0
\label{eq:consProb2a}
\end{eqnarray}
and
\begin{eqnarray}
\partial_{t_{2}} \int_{a}^{b} PDF[x_{1},t_{1},x_{2},t_{2}] dx_{2} +\notag \\ j_{2}[x_{1},t_{1},b,t_{2}]-j_{2}[x_{1},t_{1},a,t_{2}]=0.
\label{eq:consProb2b}
\end{eqnarray}
These equations indicate that the time rate of change in probability within the $\overline{ab}$ segment on the x-axis is determined separately by a net change in particle and well or barrier probability fluxes in that region, which is similar to conservation of probability in a one-body system. Now, however, probability of the two-body system is conserved even if the particle and well or barrier are measured at different times.

The incident and reflected particle-well or barrier states interfere when they overlap. This is similar to the one-body standing wave interference of a harmonic electromagnetic wave reflecting from a stationary mirror \cite{Wiener}. Classically, the incident and reflected waves interfere while the mirror experiences a continuous force due to radiation pressure. Quantum mechanically, interference occurs since the incident and reflected states are indistinguishable for a measurement of position (but not for a momentum measurement which distinguishes direction).

The two-body quantum analogy involves solving the Schr\"odinger equation using particle and well or barrier coordinates with the appropriate interaction potential. Interference is expected between the incident and reflected particle substates {\em along with} interference of the well or barrier substates which have and have not reflected the particle. Their correlation is perhaps not, being a consequence of the solution to the Schr\"odinger equation, from which a joint PDF is constructed. This then describes the correlations in the two-body interference which are manifest as coincidence rates, e.g. a correlation in the measurement of particle-well or barrier positions.

\subsection{Asynchronous model}
\label{sec:non-simultaneous}

Before reflection, the Schr\"odinger equation for the {\em non-interacting} particle-well or barrier state is
\begin{eqnarray}
(\hbar \partial_{x_{1}}^{2}/2m+\hbar \partial_{x_{2}}^{2}/2M
+i\partial_{t_{1}}+i\partial_{t_{2}})\Psi=0 \notag.
\label{eq:ScheqnIN}
\end{eqnarray}
As described above, this notation provides a label to which the energy and wavevector of each object is associated, which is illustrated in the solution given by
\begin{eqnarray}
\Psi_{0} \propto \exp[i (k x_{1}-\frac{\hbar k^{2}}{2m}t_{1}+K x_{2}-\frac{\hbar K^{2}}{2M}t_{2})],
\label{eq:ScheqnUnentangled}
\end{eqnarray}
where $k$ and $K$ are the wavevectors for the particle and well or barrier respectively, $k=m v/\hbar$ and $K=M V/\hbar$. A wavegroup constructed from the separable particle-well or particle-barrier state in equation \ref{eq:ScheqnUnentangled} then leads to uncorrelated predictions about the probability of finding the particle at $(x_{1},t_{1})$ and well or barrier at $(x_{2},t_{2})$.

The particle-well or particle-barrier interaction is determined from the Schr\"odinger equation given by
\begin{eqnarray}
(\hbar \partial_{x_{1}}^{2}/2m+\hbar \partial_{x_{2}}^{2}/2M+PE[x_{1}-x_{2}] \notag \\
+i\partial_{t_{1}}+i\partial_{t_{2}})\Psi[x_{1},x_{2},t_{1},t_{2}]=0,
\label{eq:Scheqn}
\end{eqnarray}
where $PE[x_{1}-x_{2}]$is the potential energy associated with either the well or barrier. Of interest is a solution which yields an energy eigenstate for the particle-well or barrier interaction, for which neither the particle nor the well or barrier is localized.

A separable solution to this two-body Schr\"odinger equation results from a transformation to the center of mass and relative coordinates $x_{cm}$ and $x_{rel}$ of the system with respective energies $E_{cm}$ and $E_{rel}$ (not to be confused with the cm of the particle or cm of the well or barrier). This transformation does not change the total energy, $(\hbar K)^{2}/(2M)+(\hbar k)^{2}/(2m)=E_{rel}+ E_{cm}$. Relative and center of mass times $t_{rel}$ and $t_{cm}$ are introduced and associated with the relative and center of mass energies. These time variables satisfy the same properties as do $t_{1}$ and $t_{2}$ but in this case provide the notation needed in separating the energies associated with the relative and center of mass subsystems.

The transformed Schr\"odinger equation becomes
\begin{eqnarray}
(\frac{\hbar \partial_{x_{cm}}^{2}}{2M_{tot}}+\frac{\hbar \partial_{x_{rel}}^{2}}{2 \mu}+PE[x_{rel}] \notag 
+i\partial_{t_{cm}} \\ +i\partial_{t_{rel}}) \Psi[x_{cm},x_{rel},t_{cm},t_{rel}]=0\notag ,
\label{eq:ScheqnCM}
\end{eqnarray}
where $M_{tot}=m+M$, $\mu=mM/(m+M)$, $x_{cm}=(mx_{1}+Mx_{2})/M_{tot}$, and $x_{rel}=x_{1}-x_{2}$. Using 
\begin{eqnarray}
\Psi[x_{cm},x_{rel},t_{cm},t_{rel}]=\psi_{cm}\psi_{rel}  \notag \\ =e^{-i E_{cm} t_{cm}/\hbar} U[x_{cm}] e^{-i E_{rel} t_{rel}/\hbar}u[x_{rel}] \notag ,
\label{eq:Schtot}
\end{eqnarray}
reduces the Schr\"odinger equation to two ordinary differential equations:
\begin{equation}
-\frac{\hbar}{2M_{tot}} \frac{d^{2}U[x_{cm}]}{dx_{cm}^{2}} = E_{cm} U[x_{cm}] 
\label{eq:ScheqODE1}
\end{equation}
\begin{equation}
-\frac{\hbar}{2 \mu} \frac{d^{2}u[x_{rel}]}{dx_{rel}^{2}}+PE[x_{rel}] = E_{rel} u[x_{rel}].
\label{eq:ScheqODE2}
\end{equation}

\subsection{Infinite-potential well energy eigenstates}
\label{sec:well}

The boundary condition for the infinite potential well is $\Psi[x_{rel} \pm D]\rightarrow 0$. The relative wavefunction does not exist outside the well for $x_{rel}<-D$ and $x_{rel}>D$. A solution to eqn. \ref{eq:ScheqODE2} for a particle which has reduced mass is \cite{serge}
\begin{equation}
\Psi_{rel}  \propto e ^{-i E_{rel} t_{rel}/\hbar } \sin[\frac{n \pi (x_{rel}+D)}{2D}]~\Theta[x_{rel},D],
\label{eq:Psiwell2}
\end{equation}
where $\Theta[x_{rel},D]=(\theta [x_{rel}+D]-\theta [x_{rel}-D])$ ($\theta$ is the unit step function) has value one within the well and zero everywhere else with n being the number of nodes.

The solution to eqn. \ref{eq:ScheqODE1} is given by 
\begin{equation}
\Psi_{cm} \propto e ^{i (K_{cm} x_{cm}-E_{cm} t_{cm}/\hbar }).
\label{eq:Psiwell1}
\end{equation}
The complete solution is then $\Psi[x_{cm},x_{rel},t_{cm},t_{rel}] \propto \psi_{cm}\psi_{rel}$.

We are interested in measurements of the particle and well rather than measurements of two transformed ``objects,'' one with a reduced mass and the other with the total mass of the system, neither of which exist. A transformation from the relative and center of mass coordinates (not to be confused with the cm of the particle and well) to the particle-well coordinates is then needed and involves the following substitutions: $K_{cm}=k+K$, $K_{rel}=(Mk-mK)/M_{tot}$, $x_{rel}=x_{2}-x_{1}$, $x_{cm}=(mx_{1}+Mx_{2})/M_{tot}$, $E_{rel}=\hbar^2K_{rel}^{2}/2\mu$, and $E_{cm}=\hbar^2K_{cm}^{2}/2(m+M)$. 

Expanding the sine function in eqn. \ref{eq:Psiwell2} into exponential form results in wavefunctions traveling in opposite directions. Transforming back to the particle-well coordinates, the momenta and energies of the particle and well differ in magnitude in these two directions. 

The modes of the square well, characterized by the number of nodes $n$, are determined in the cm-rel coordinates by $K_{rel}=n\pi/2D$. In the particle-well coordinates this constrains values of particle and well velocities by
\begin{equation}
v=V+n \pi \frac{\hbar (m+ M)}{2DmM}.
\label{eq:quantizationvV}
\end{equation}

The wavefunction's phase, $\phi$, needs to have its temporal part transformed from the parameters $t_{cm}$ and $t_{rel}$ to $t_{1}$ and $t_{2}$. This is done by associating the kinetic energy of the particle with $t_{1}$ and the kinetic energy of the well with $t_{2}$. The particle's kinetic energy, $p_{particle}^{2}/2m$ is determined from $p_{particle}=\hbar \partial \phi/\partial x_{1}$. Similarly, the energy for the well, $p_{well}^{2}/2M$, is determined from $p_{well}=\hbar \partial \phi/\partial x_{2}$. These values differ in the two directions. While the resulting energies and momenta are consistent with those of a classical particle reflecting from the walls of a moving well they are manifest in the two-body wavefunctions as a ``Doppler shift''. The resulting expression for the eigenstate of energy, $\Psi[x_{1},x_{2},t_{1},t_{2}]$, is too large to present here.

\subsection{Potential barrier energy eigenstates}
\label{sec:barrier}

The barrier has potential energy $PE$ and extends over a distance $D$. This divides space into three regions: before the barrier or ``before'', in the barrier region or ``barrier,'' and after the barrier or ``after.'' Solutions are first obtained for these three regions in the cm and rel coordinates by solving eqns. \ref{eq:ScheqODE1} and \ref{eq:ScheqODE2}.

The solution to eqn. \ref{eq:ScheqODE2} before the barrier consists of incident and reflected wavefunctions given by
\begin{eqnarray}
\Psi_{rel}^{before} = A e^{i (K_{before} x_{rel}-E_{rel} t_{rel}/\hbar }) \notag \\ +B e^{i (-K_{before} x_{rel}-E_{rel} t_{rel}/\hbar }),
\label{eq:Psibefore}
\end{eqnarray}
where $K_{before}=\sqrt{2\mu E_{rel}}/\hbar$. The solution in the barrier region also consists of incident and reflected wavefunctions given by
\begin{eqnarray}
\Psi_{rel}^{barrier} = F e^{i (K_{barrier} x_{rel}-E_{rel} t_{rel}/\hbar }) \notag \\ +G e^{i (-K_{barrier} x_{rel}-E_{rel} t_{rel}/\hbar }),
\label{eq:Psibarrier}
\end{eqnarray}
where $K_{barrier}=\sqrt{2\mu (E_{rel}-PE)}/\hbar$. The solution after the barrier consists only of a transmitted wavefunction given by
\begin{eqnarray}
\Psi_{rel}^{after} = H e^{i (K_{after} x_{rel}-E_{rel} t_{rel}/\hbar }),
\label{eq:Psiafter}
\end{eqnarray}
where $K_{after}=\sqrt{2\mu E_{rel}}/\hbar$.

The boundary conditions are continuity of the wavefunctions and their derivatives with respect to $x_{rel}$ at $x_{rel}=\pm D$. These then constrain the coefficients $B$, $F$, $G$, and $H$ with $A=1$. 

Again, the solution to eqn. \ref{eq:ScheqODE1} is given by eqn. \ref{eq:Psiwell1}. The complete solution is then $\Psi[x_{cm},x_{rel},t_{cm},t_{rel}] \propto \psi_{cm}\psi_{rel}$. Since we are interested in predictions about measurements of the particle and well rather than measurements of the reduced mass and total mass ``objects,'' a transformation from the relative and center of mass to the particle-well coordinates is utilized.

The wavefunction's phase, $\phi$, needs to have its temporal part transformed from the parameters $t_{cm}$ and $t_{rel}$ to $t_{1}$ and $t_{2}$. This is done by associating the kinetic energy of the particle with $t_{1}$ and the kinetic energy of the well with $t_{2}$. The particle's kinetic energy, $p_{particle}^{2}/2m$ is determined from $p_{particle}=\hbar \partial \phi/\partial x_{1}$. Similarly, the energy for the barrier, $p_{barrier}^{2}/2M$, is determined from $p_{barrier}=\hbar \partial \phi/\partial x_{2}$. This has to be calculated separately in each region and wave propagation direction. The resulting energies and momenta are consistent with those of a classical particle reflecting from a barrier and are manifest in the two-body wavefunctions as a ``Doppler shift''. Again, the resulting expression for the eigenstate of energy, $\Psi[x_{1},x_{2},t_{1},t_{2}]$, is too large to present here.

Since this procedure does not assign a time variable to the potential energy in the barrier region, a time $t_{PE}$ is used for this purpose. Such a term has no measurable effect as will be shown next. The wavefunction in the barrier, expressed in terms of the particle-well coordinates, is then 
\begin{eqnarray}
\Psi^{barrier} = F e^{i \{\Phi_{spatial}^{right}-(KE_{1}^{right} t_{1}+KE_{2}^{right} t_{2}+PE~t_{PE})/\hbar\}} \notag \\
 +G e^{i \{\Phi_{spatial}^{left}-(KE_{1}^{left} t_{1}+KE_{2}^{left} t_{2}+PE~t_{PE})/\hbar\}},\notag
\label{eq:Psibarrierlab}
\end{eqnarray}
where $\Phi_{spatial}^{right}$ and $\Phi_{spatial}^{left}$ contain the spatial terms in the phase and are functions of $m,M,v,V,PE,x_{1},x_{2},$ and $\hbar$. The temporal terms contain the kinetic energy for the particle and barrier moving to the right and left, $KE_{1}^{right},KE_{2}^{right},KE_{1}^{left},KE_{2}^{left}$ and the potential energy $PE$. The kinetic energy terms are functions of $m,M,v,V,PE$, and $\hbar$. 

The potential energy term $e^{i PE~t_{PE}}/\hbar$ is a common factor of both the incident and reflected wavefunctions in the barrier. Since the PDF is generated from the wavefunction multiplied by its complex conjugate, such common factors have no effect on the PDF. There is then  no need to associate the potential energy part of the total energy with either the particle, as a coefficient of $t_{1}$, or the barrier, as a coefficient of $t_{2}$, or as a separate term in the phase, $PE~t_{PE}/\hbar$. The potential energy part of the total energy has observable consequences only parametrically within the momenta and kinetic energies of the particle and barrier. This simple division of the momentum and energy of the particle in the barrier can be contrasted with that of the stress-energy tensor for an electromagnetic wave traversing a dielectric slab \cite{loudon}. 

\section{Wavegroup results}
\label{sec:wavegroup}

Wavegroups are next formed from a Gaussian superposition of the two-body energy eigenstates for the particle and well or barrier described above. Correlated interference is a consequence of such a superposition. However, we focus the following discussion on the subsets of correlated interference effects which deal with the superposition of two such wavegroups. One example is the interference when the incident and `reflected' two-body wavegroups overlap. These types of PDF's are first illustrated for {\em simultaneous} measurement of the particle and well or barrier, followed by ones for {\em asynchronous} measurements.  

\subsection{Particle-barrier or finite well}
\label{sec:barriersimultaneous}

The wavegroup for the particle and barrier or finite well is calculated using a Gaussian superposition of the energy eigenstates given in subsection \ref{sec:barrier}. Unfortunately, the coefficients $B$, $F$, $G$, and $H$ of eqns. \ref{eq:Psibefore}, \ref{eq:Psibarrier}, and \ref{eq:Psiafter} depend in a non-trivial manner on the variables of integration. The resulting integrals cannot be determined in closed form. To facilitate the calculation, the following sums will replace these integrals:
\begin{eqnarray}
\Psi_{barrier}^{wavegroup} \propto  \sum_{V_{i}}^{V_{f}} \frac{e^{\frac{-(V-V_{0})^{2}}{2\Delta V^{2}}}}{\sqrt{\Delta V}} \Psi[x_{1},x_{2},t_{1},t_{2}] \notag
\label{eq:barrierwavegroup}
\end{eqnarray}
where the peak of the barrier velocity distribution is at $V_{0}$, $\Delta V$ is its width, and the sum is from an initial barrier velocity $V_{i}$ to a final velocity $V_{f}$. Summing over the particle velocity distribution yields the wavefunction for the wavegroup given by
\begin{eqnarray}
\Psi_{total}^{wavegroup} \propto  \sum_{v_{i}}^{v_{f}} \frac{e^{\frac{-(v-v_{0})^{2}}{2\Delta v^{2}}}}{\sqrt{\Delta v}} \Psi_{barrier}^{wavegroup}, \notag
\label{eq:totbarrierwavegroup}
\end{eqnarray}
where the peak of the particle velocity distribution is at $v_{0}$, $\Delta v$ is its width, and the sum is from an initial particle velocity $v_{i}$ to a final velocity $v_{f}$.

\subsubsection{Simultaneous measurement: $KE_{initial}^{relative}>PE$}
\label{sec:barriergreater}

\begin{center}
\begin{figure}
\includegraphics[scale=0.29]{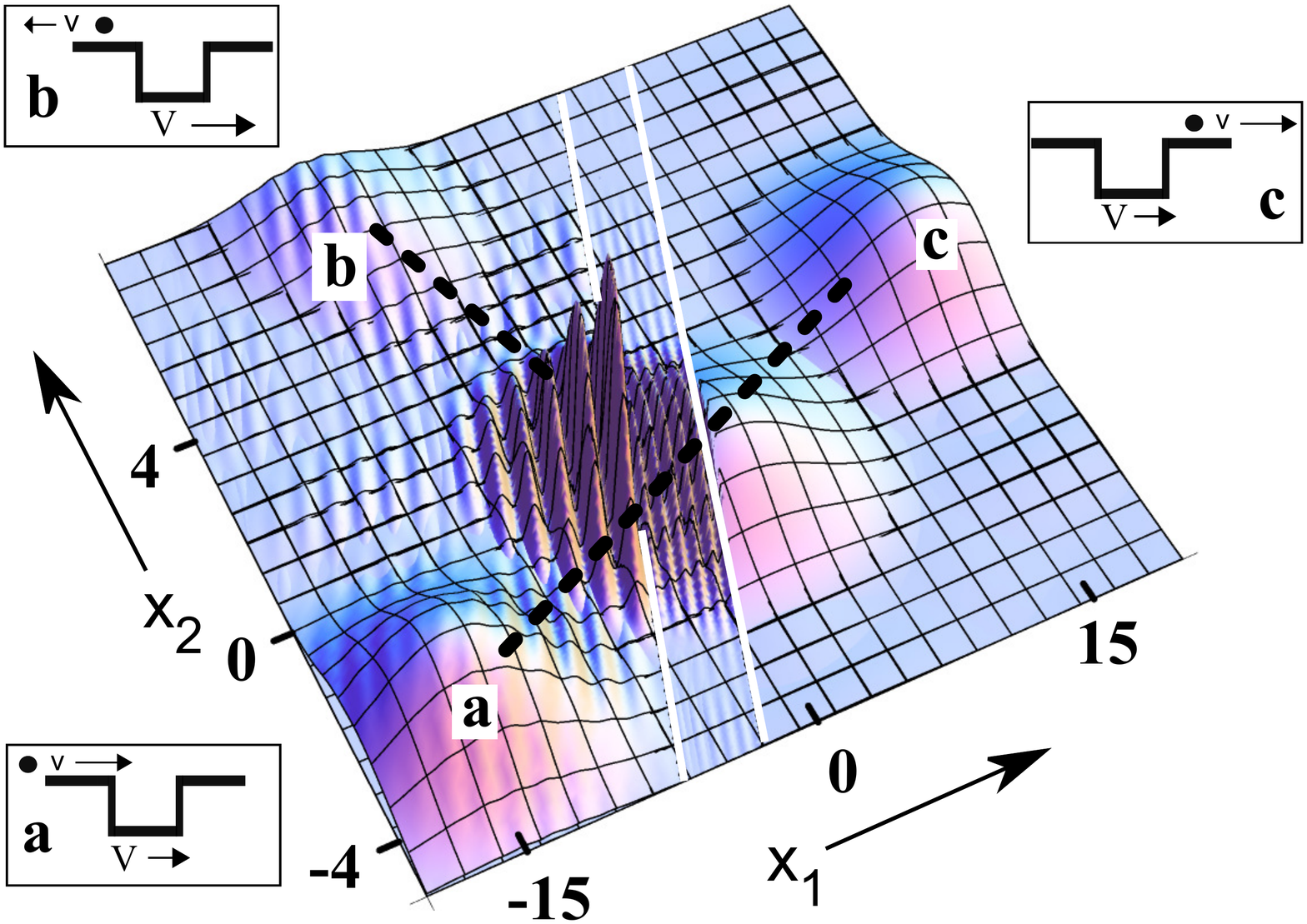}
\caption{Three {\em simultaneous} PDF snapshots for sequential times vs coordinates $(x_{2},x_{1})$. The first snapshot generates peak $a$ while peaks $b$ and $c$ comprise the last snapshot. The classical analogs of these peaks are shown in the insets. The PDF progresses temporally along the dashed line.}
\label{fig:barrier_negenergy}
\end{figure}
\end{center}

Consider next a particle interacting with a finite well. Their sum of initial kinetic energies in the relative coordinate system is greater than the well PE. The size of the particle substate wavegroup is chosen to be a few times larger than the finite well width $D$. Fig. \ref{fig:barrier_negenergy} shows results of the PDF's for three sequential snapshots taken at equal time intervals progressing along the dashed line from the lower left to upper right and upper left. The analagous classical positions of the particle and finite well for particular snapshots are illustrated in the insets. The barrier boundaries occur at the diagonal white lines, corresponding to $x_{1}=x_{2} \pm D$. The parameters used in fig. \ref{fig:barrier_negenergy} are $v_{0}/V_{0}=6$, $\Delta v/\Delta V=1.5$, and $M/m=5$ while the $KE_{initial}^{relative}-PE/\mid PE\mid =1.4$ using the average value of $KE_{initial}^{relative}$ for the wavegroup particle and well distributions. 

One category of correlated interference, which we call type I, occurs when the incident and reflected two-body wavefunctions, traveling in opposite directions in the $(x_{2},x_{1})$ plane, `overlap.' This is illustrated in fig. \ref{fig:barrier_negenergy} by the fringes of the middle snapshot just to the left of the line $x_{2}=x_{1} + D$. These fringes are spaced by about half the deBroglie wavelength of the {\em particle} for $M>>m$ and $v>>V$ as are the similar correlated interference fringes in the two-body reflection of a particle from a mirror \cite{kowalski}. 

However, the interaction generates another form of correlated interference when the reflected wavegroups, one from each barrier surface, interfere as they travel in the {\em same} direction in the $(x_{2},x_{1})$ plane. This new category of correlated interference, which is referred to as type II, is illustrated in fig. \ref{fig:barrier_negenergy} by the peak labeled $b$ and is similar to the classical interference of a pulse of light reflecting from a thin film. 

Changing only the barrier spacing generates an oscillation in the PDF for peak $b$ analogous to that found in the interference of a pulse of light reflecting from a thin film whose thickness varies. That is, this peak goes through constructive and destructive interference from the two barrier reflections when the wavegroup size is much larger than that of the barrier and the spacing $D$ is varied. As time progresses peak $b$ maintains this interference as it travels in the $(x_{2},x_{1})$ plane, whereas the correlated interference associated with the fringes shown in the middle snapshot of fig. \ref{fig:barrier_negenergy} is localized to a small temporal and spatial region.

\subsubsection{Simultaneous measurement: $PE < KE_{initial}^{relative} < PE$}
\label{sec:barriermiddle}

Consider next wavegroups for which some Fourier components of the particle and barrier substates have a total initial kinetic energy in the relative coordinate system which exceeds the barrier potential energy while other components have a total relative initial kinetic energy which is less than the barrier potential. To illustrate the resulting PDF's the size of the particle and barrier substate wavegroups are chosen to be slightly larger than the barrier width $D$. 

\begin{center}
\begin{figure}
\includegraphics[scale=0.29]{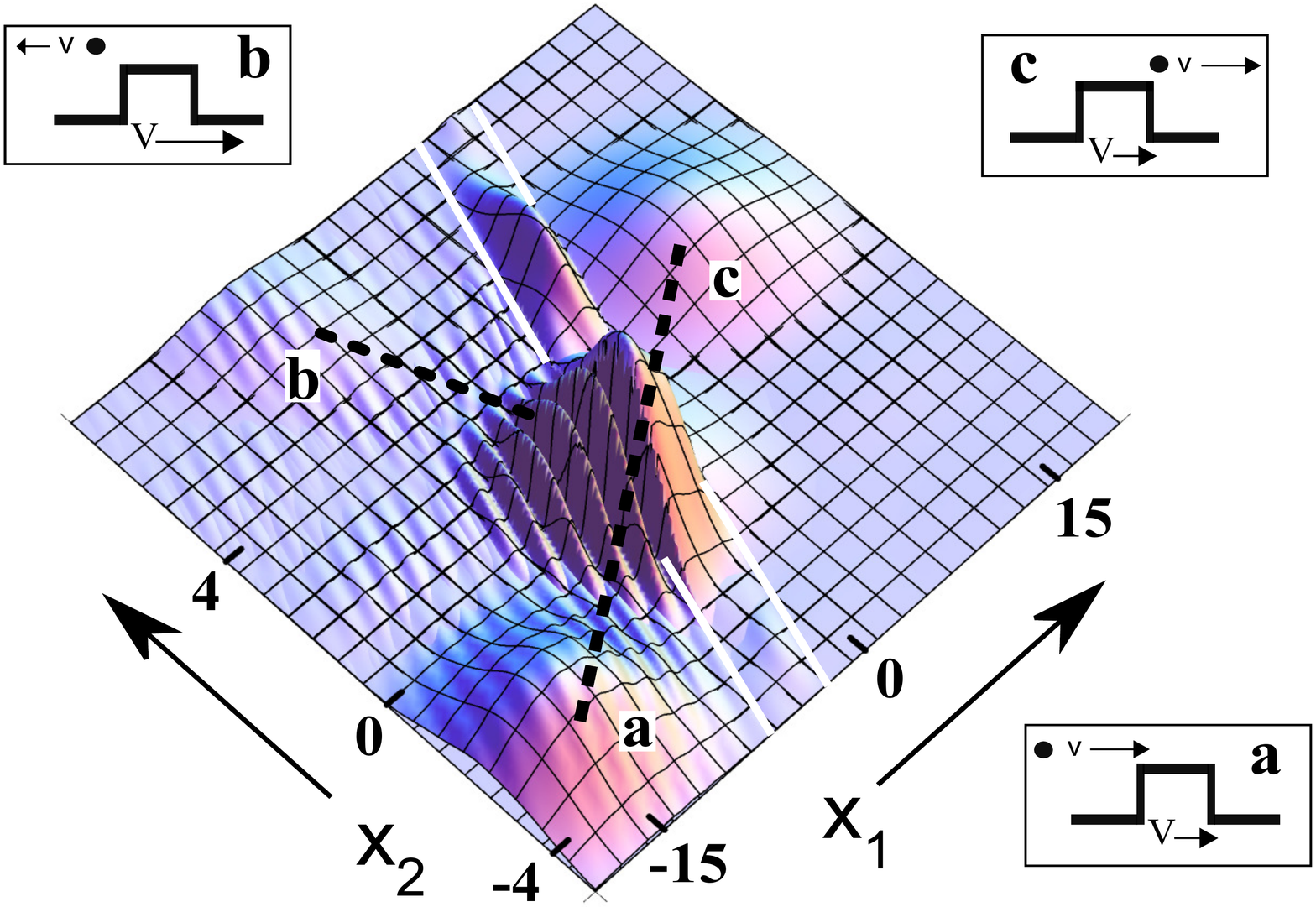}
\caption{Three {\em simultaneous} PDF snapshots for sequential times vs coordinates $(x_{2},x_{1})$ for the particle traversing the barrier. The only difference between the parameters used here and in fig. \ref{fig:barrier_negenergy} is the PE which forms a barrier.}
\label{fig:barrier_posenergy}
\end{figure}
\end{center}

Fig. \ref{fig:barrier_posenergy} shows the PDF's from such an interaction using three sequential snapshots, progressing along the dashed line from the lower left to upper right. The speed of the particle and well are illustrated for a classical system of two such particles in the insets next to each snapshot. Again the diagonal white lines correspond to $x_{1}=x_{2} \pm D$. The parameters used in fig. \ref{fig:barrier_posenergy} are $v_{0}/V_{0}=6$, $\Delta v/\Delta V=1.5$, $M/m=5$ while the $KE_{initial}^{relative}-PE/\mid PE\mid =0.3$ for the average value of the $KE_{initial}^{relative}$ for the wavegroup distributions. 

This figure illustrates yet another form of correlated interference, referred to as type III: that from multiple reflections from the two barrier edges, which is shown isolated from other effects as the unlabeled peak in the PDF located between both $x_{1}=x_{2} \pm D$ and peaks $b$ and $c$ in the third snapshot. This peak is analogous to the buildup and decay of electromagnetic energy in a optical cavity. Later snapshots (not shown) illustrate its decay.

The position of peak $c$ can be compared between figs. \ref{fig:barrier_negenergy} and \ref{fig:barrier_posenergy} since all parameters are the same except the PE. The location of this peak indicates the effect of the interaction on the relative transit times for the particle and finite well or barrier wavegroup substates.

\subsubsection{Asynchronous results: particle-barrier $KE_{initial}^{relative}>PE$}
\label{sec:barrierasynchronous}

Next consider asynchronous measurements of the particle at a particular $x_{1}$ and $t_{1}$ while the barrier substate then evolves with $\Psi[x_{1}|_{fixed},x_{2},t_{1}|_{fixed},t_{2}]$. Once the particle has been measured there is no interaction between the particle and barrier. The sum of the relative initial kinetic energies is chosen to always be greater than the barrier PE. Asynchronous predictions are described first for type II and then type I correlated interference. As an example of the former we fix $x_{1}$ and $t_{1}$ on peak (b) in fig.\ref{fig:barrier_negenergy} while for the latter $x_{1}=-2$ and $t_{1}$ is fixed during the middle snapshot time shown in fig. \ref{fig:barrier_negenergy}.

\begin{center}
\begin{figure}
\includegraphics[scale=0.29]{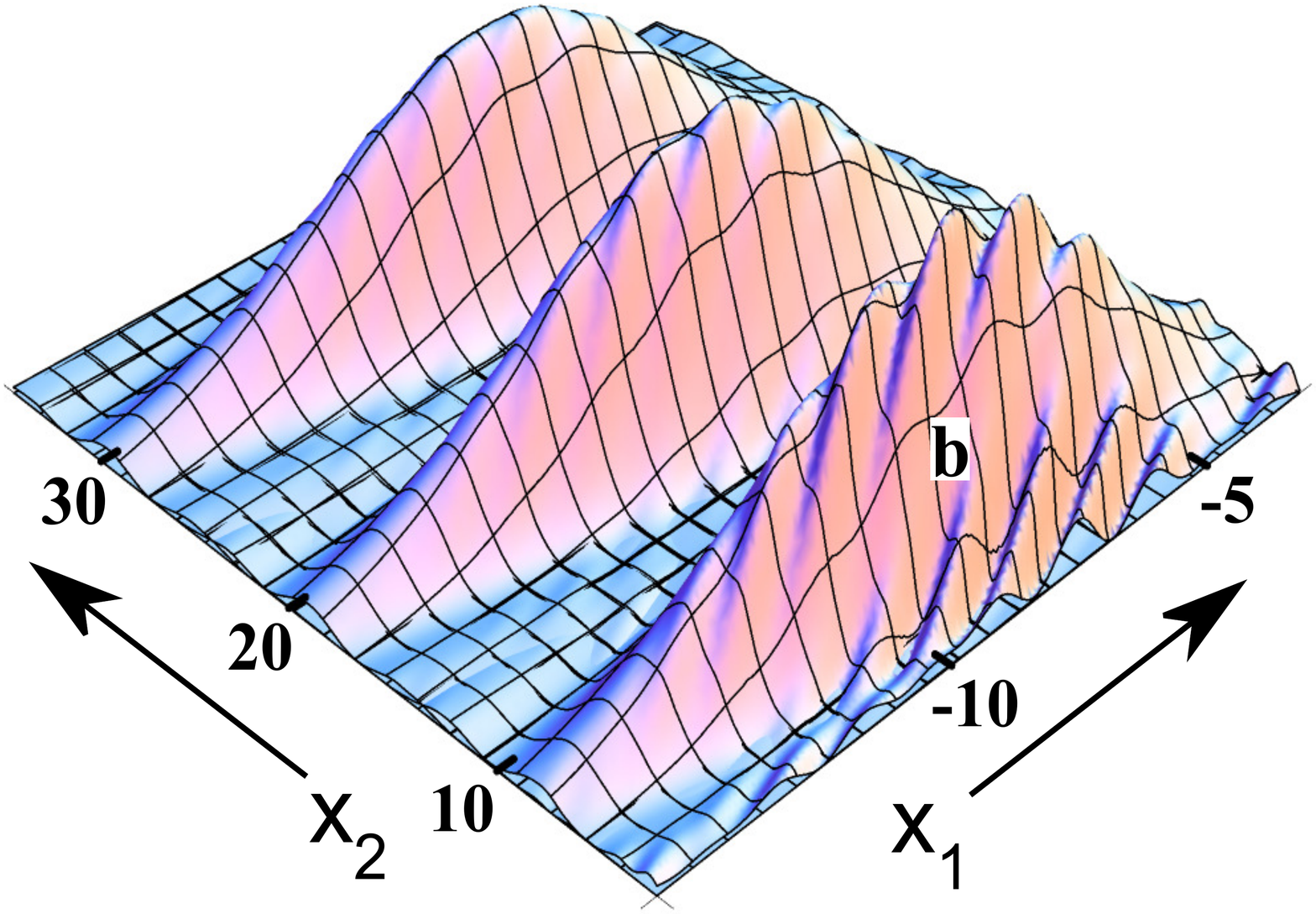}
\caption{{\em Asynchronous} PDF snapshots for equal sequential times $t_{2}$ when the particle is measured during type II correlated interference. 
The snapshot labeled (b) is for $t_{2}=t_{1}$ and is therefore the same as peak (b) in fig. \ref{fig:barrier_negenergy}. }
\label{fig:barrierAsynchPeakB}
\end{figure}
\end{center}

The most straightforward way to illustrate asynchronous measurement in type II correlated interference is to graph the barrier's PDF vs. $x_{2}$ for different snapshots in $t_{2}$ while fixing $x_{1}$ and $t_{1}$ on peak (b) in fig. \ref{fig:barrier_negenergy}. This is shown in fig. \ref{fig:barrierAsynchPeakB}. The lowest snapshot, labeled (b), is for $t_{2}=t_{1}$ and is therefore the same as peak (b) in fig. \ref{fig:barrier_negenergy} while the other snapshots sequentially increase only $t_{2}$.

\begin{center}
\begin{figure}
\includegraphics[scale=0.29]{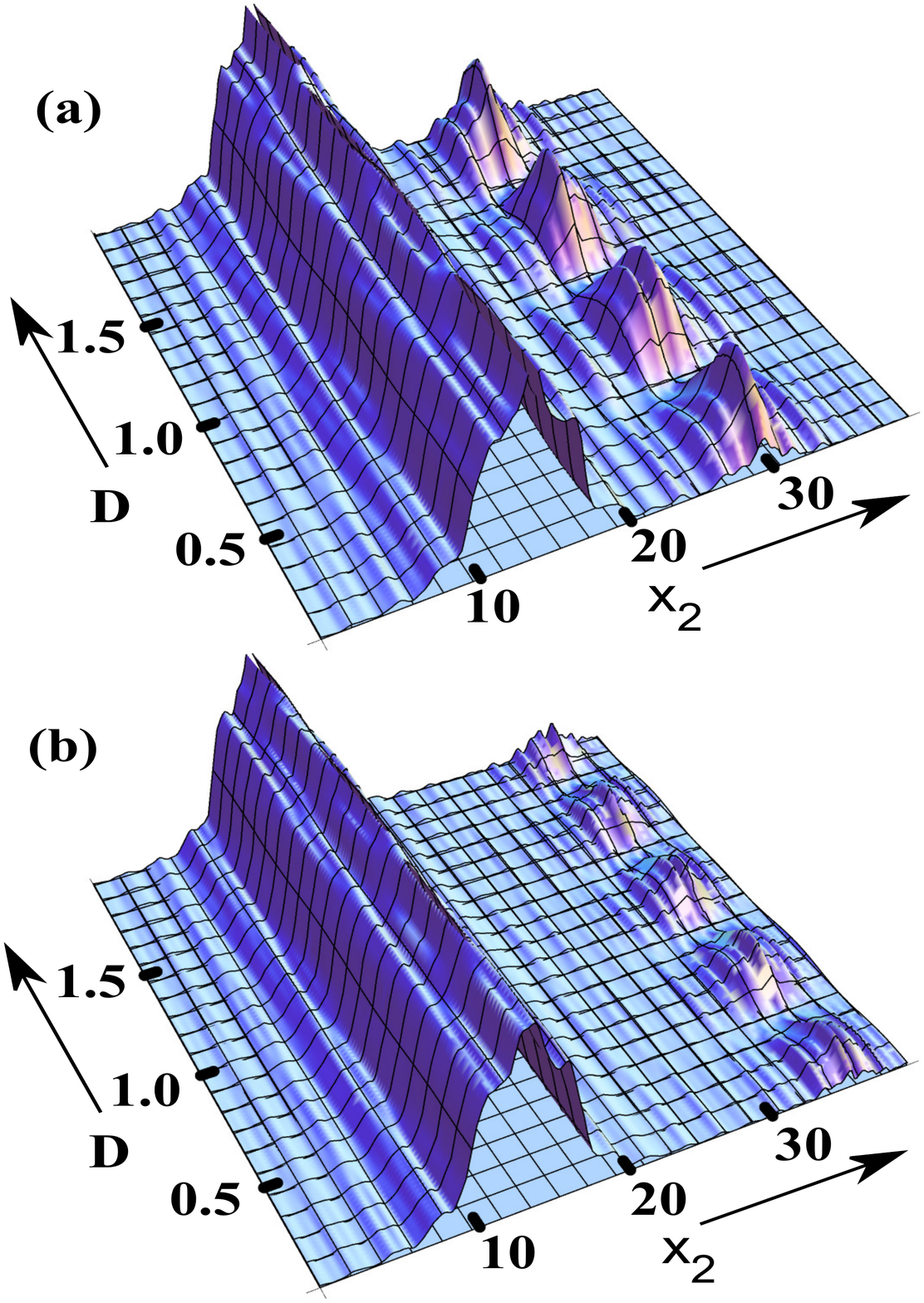}
\caption{{\em Asynchronous} PDF snapshots when the particle is measured during type I correlated interference. The two peaks indicate a splitting of the barrier substate due to measurement of the particle at an earlier time.}
\label{fig:barrierAsynchfinal}
\end{figure}
\end{center}

To illustrate asynchronous measurement in type I correlated interference, two 3-D plots of the barrier's PDF vs. $x_{2}$ {\em and} $D$ are shown in fig. \ref{fig:barrierAsynchfinal} while fixing $x_{1}$ and $t_{1}$ to be in the space-time region associated with the type I correlated inteference of the middle snapshot in fig. \ref{fig:barrier_negenergy} ($x_{1}=-2$ and $t_{1}$ is time associated with this snapshot). While both plots are for the same time $t_{2}$, fig. \ref{fig:barrierAsynchfinal} (a) is for a smaller initial particle velocity than is fig. \ref{fig:barrierAsynchfinal} (b). 

Since the particle is measured in the space-time region associated with type I correlated inteference, the particle could have come from the incident or reflected wavegroups which travel in opposite directions. The largest peak in fig. \ref{fig:barrierAsynchfinal} corresponds to the barrier substate when the measured particle did not interacted with the barrier (the measured particle came from the incident wavegroup moving to the right). The position of this peak is then the same for different particle speeds as is shown in fig. \ref{fig:barrierAsynchfinal} and is consistent with the expected position of the barrer for no interaction.

The smaller peak corresponds to the particle having been measured after it reflected (the measured particle came from the reflected wavegroup moving to the left in the middle snapshot of fig. \ref{fig:barrier_negenergy}). The position of this smaller barrier PDF peak therefore increases with increased particle speed due to recoil of the barrier, as shown by the different $x_{2}$ positions of the smaller peaks in fig. \ref{fig:barrierAsynchfinal} (a) and (b). These positions are consistent with those expected from ``classical recoil'' of the barrier from the particle moving at these different speeds. This ``classical recoil'' corresponds to the particle reflecting only once from the barrier even though the wave reflects from both the front and back edges of the barriers. The standard interpretation of such interference is that there is an amplitude for the particle to reflect from the front and an amplitude to refelct from the back barrier edge. These two amplitudes then interfere.

Since the particle could have reflected from either barrier edge, the barrier is in a superposition of both these possibilities. The PDF as a function of barrier width along the $D$ axis illustrates interference due to the barrier being put into a superposition in which the particle either reflected from the front or rear edges of the barrier. For larger particle speeds the wavelength decreases, resulting in the increased number of fringes along the $D$ axis as shown in this figure.

The parameters used in fig. \ref{fig:barrierAsynchfinal} (a) are  $v_{0}/V_{0}=4.6$, $\Delta v/\Delta V=1.5$, $M/m=5$ while in fig. \ref{fig:barrierAsynchfinal} (b) they only differ by the increased particle speed $v_{0}/V_{0}=6.1$. Differences in the PDF sizes and shapes of these figs. are associated with the states being expressed as sums, the transmissions and reflection coefficients depending on velocity, and the dependence of the type I fringe pattern on velocity (the PDF values for a given $x_{1}$ and $t_{1}$ depend on the particle speed).

\subsection{Infinite well-particle wavegroup}
\label{sec:wellwavegroup}

This calculation differs from that of the barrier due both to the particle and barrier velocities being contrained by the resonance condition given in eqn. \ref{eq:quantizationvV} and by the lack of coefficients, such as the $B$, $F$, $G$, and $H$ used in the previous section, which depend on the parameters of integration.

A closed form expression for the well substate wavegroup can be obtained from a Gaussian distribution of the energy eigenstates parametrized in terms of velocity components, $V$, of the well, given by
\begin{eqnarray}
\Psi_{well}^{wavegroup} \propto \int_{-\infty}^{\infty} \frac{e^{\frac{-(V-V_{0})^{2}}{2\Delta V^{2}}}}{\sqrt{\Delta V}} \Psi[x_{1},x_{2},t_{1},t_{2}]dV  \notag
\label{eq:Psiwellwavegroup}
\end{eqnarray}
where the peak of the distribution is at $V_{0}$ and $\Delta V$ is its width. This is then summed over integral values of $n$ (the number of nodes) using the gaussian distribution \cite{serge}
\begin{eqnarray}
\Psi_{total}^{wavegroup} \propto \sum_{n} \exp [-\{(n-n_{0})\pi \Delta x\}^{2}] 
\Psi_{well}^{wavegroup}, \notag 
\label{eq:Psiparticlewavegroup}
\end{eqnarray}
where the peak of the distribution is at $n_{0}$ and $\Delta x$ is its width. 

\subsubsection{Simultaneous measurement}
\label{sec:wellsimultaneous}

\begin{center}
\begin{figure}
\includegraphics[scale=0.29]{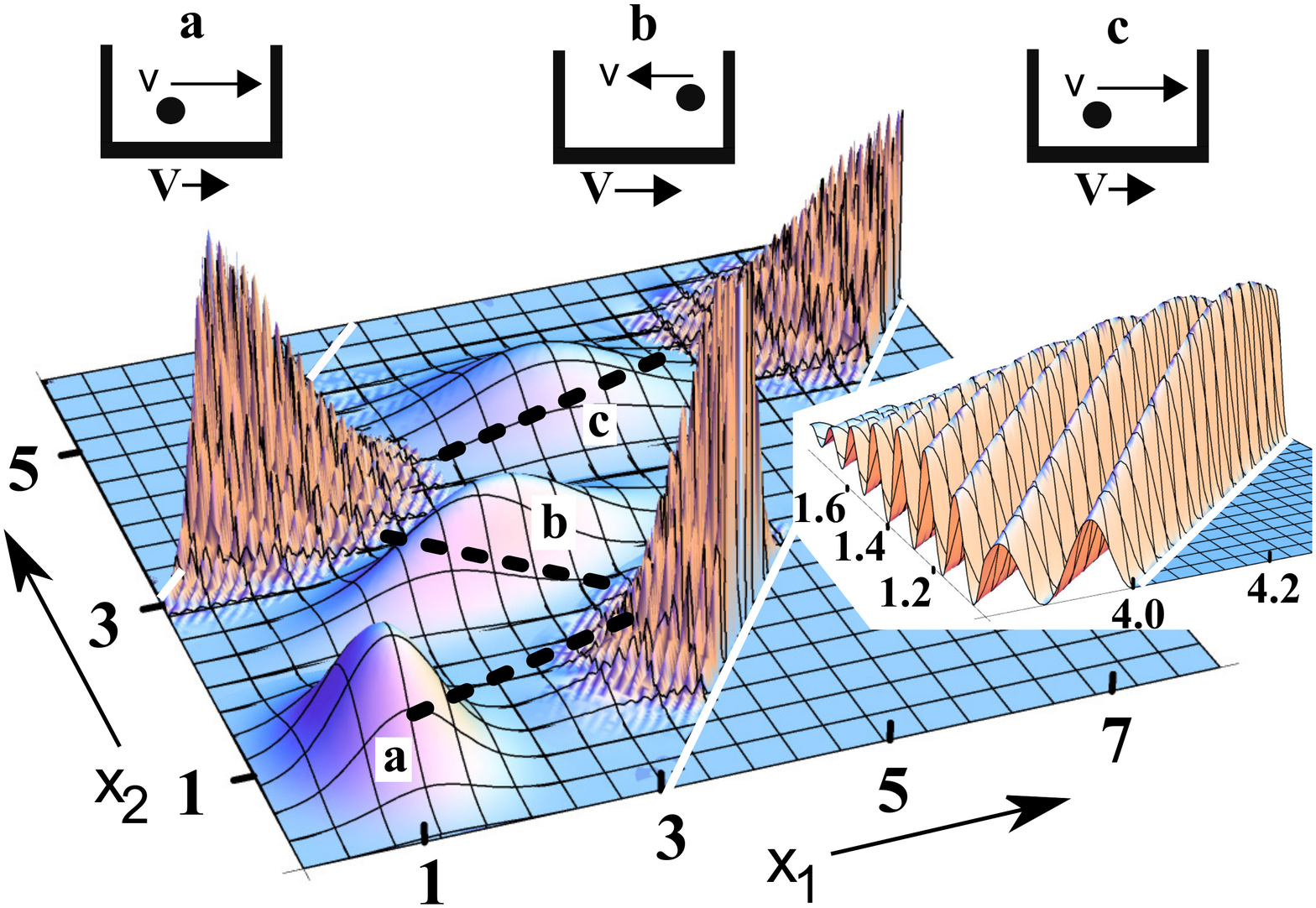}
\caption{{\em Simultaneous} PDF snapshots for sequential times vs coordinates $(x_{2},x_{1})$ for a two-body wavefunction whose wavegroup size is less than the infinite well spacing. The diagonal white lines correspond to $x_{1}=x_{2} \pm D$. The trajectory is indicated by the dashed line. The right inset is a blow-up of the correlated interference of the second snapshot. The upper insets illustrate the classical analogies for the respective wavegroup snapshots.}
\label{fig:wellsynchfinal}
\end{figure}
\end{center}

Using these relations, the PDF is plotted first with particle and well substate wavegroups whose spatial widths are less than the well spacing $D$. Fig. \ref{fig:wellsynchfinal} shows such PDF results for six snapshots taken at equal time intervals progressing from the lower left to upper right along the dashed line. `Reflection' occurs at the diagonal white lines, corresponding to $x_{1}=x_{2} \pm D$.  The particle and well ``reflect'' from each other twice, in the second and fourth snapshots. The classical analogs for the particle and well positions for some snapshots are shown in the insets, labeled by a, b, and c. These correspond to the snapshots of the wavegroups labeled with the respective letters. While the insets are schematics of the `classical' analog between the wave and particle pictures, there is nothing similar for the correlated interference snapshots. The parameters used in fig. \ref{fig:wellsynchfinal} are $n_{0}=50$, $\Delta x/D=1/15$, $\Delta V/V_{0}=1/30$, and $M/m=10$.

Type I correlated interference occurs when the incident and reflected two-body wavefunctions `overlap' and is shown in higher spatial resolution for the first reflection in the right side inset of fig. \ref{fig:wellsynchfinal}. The fringes are spaced by about half the deBroglie wavelength of the {\em particle} for $M>>m$ and $v>>V$, as are similar correlated interference fringes in two-body reflection of a particle from a mirror \cite{kowalski} 

\begin{center}
\begin{figure}
\includegraphics[scale=0.35]{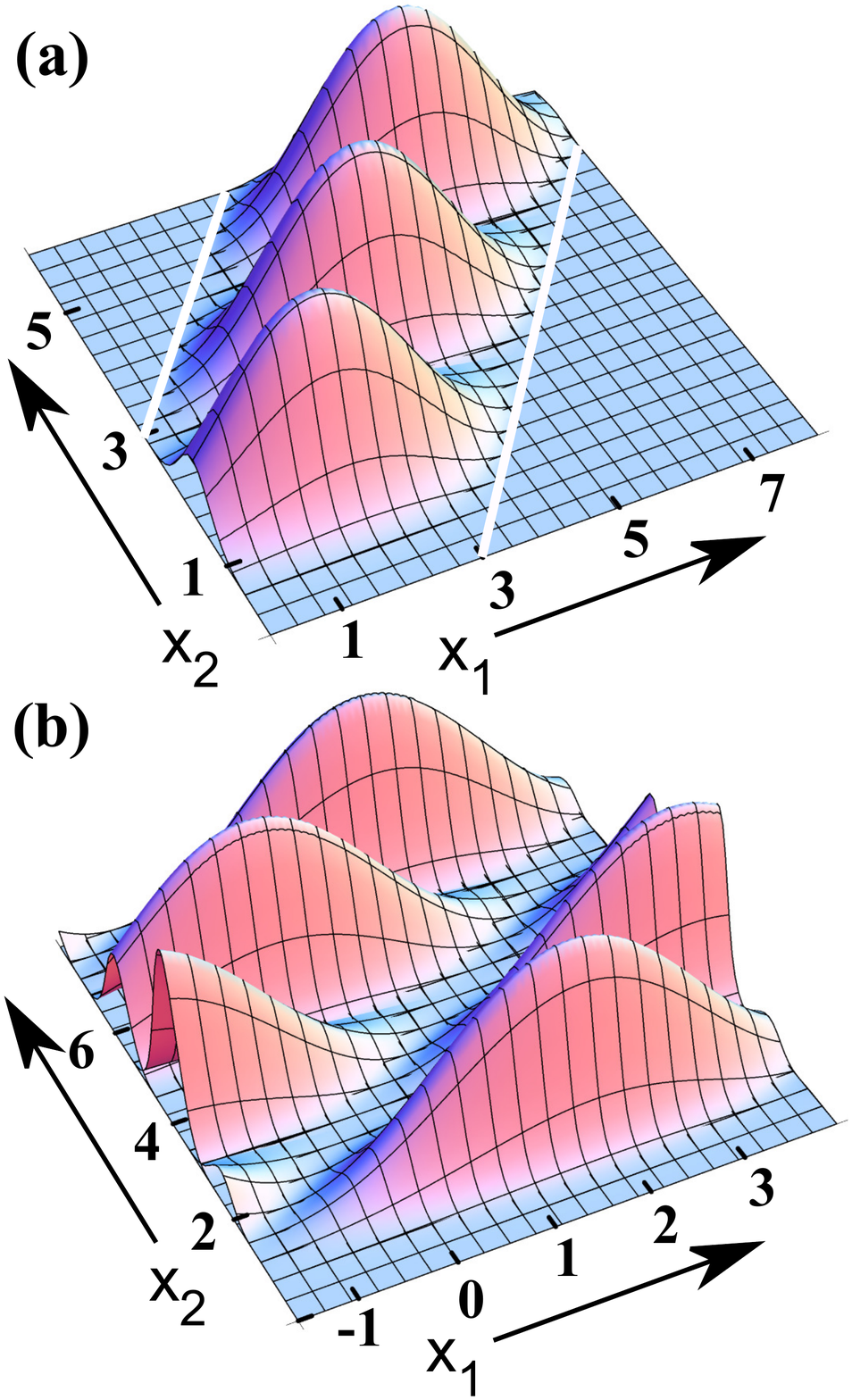}
\caption{{\em Simultaneous} (a) and {\em asynchronous} (b) PDF snapshots for sequential times vs coordinates $(x_{2},x_{1})$ for $n=1$. The upper four snapshots in (b) occur for sequential increases in $t_{2}$ while $t_{1}$ is fixed at the lowest snapshot time of (a). The lowest snapshots in (a)and (b) are therefore the same. }
\label{fig:WellStandingWave}
\end{figure}
\end{center}

Next, particle and well substate wavegroups are chosen so that the spatial width of the particle substate fills the well spacing $D$ by using only the ground state $n=1$ while the well substate width is much less than $D$.  Fig. \ref{fig:WellStandingWave} (a) shows such PDF results of three snapshots taken at equal time intervals progressing from the lower left to upper right. `Reflection' again occurs at the diagonal white line, corresponding to $x_{1}=x_{2} \pm D$. This particle substate is a superposition of waves traveling in opposite directions with different energies, resulting in the PDF appearing as `traveling wave' that matches the boundary conditions at $x_{1}=x_{2} \pm D$. The parameters used in fig. \ref{fig:WellStandingWave} are $n=1$, $\Delta x/D=1/15$, $\Delta V/V_{0}=1/30$, and $M/m=10$.

\subsubsection{Asynchronous measurement}
\label{sec:wellasynchronous}

To illustrate asynchronous measurements, the particle is measured at a particular $x_{1}$ with $t_{1}$ being the time of the lower snapshot in fig. \ref{fig:WellStandingWave} (a), while the well substate then evolves with $\Psi[x_{1}|_{fixed},x_{2},t_{1}|_{fixed},t_{2}]$. Rather than graph this time evolution using a 2-D plot of the well's PDF vs. $x_{2}$ for different $t_{2}$ snapshots and various values of $x_{1}$, PDF plots of snapshots at different times $t_{2}$ are shown on a 3-D graph of $x_{2}$ vs fixed values of $x_{1}$ as shown in fig. \ref{fig:WellStandingWave} (b). To obtain the aformentioned plot from fig. \ref{fig:WellStandingWave} (b) one must take a slice along the $x_{2}$ axis for a fixed value of $x_{1}$. Both figures \ref{fig:WellStandingWave} (a) and (b) use the same parameters. Although the time intervals per snapshot are the same for both figures, more snapshots are shown as a vertical progression in part (b). 

The resulting time dependent PDF along the $x_{2}$ axis for fixed $x_{1}$ and $t_{1}$ is a consequence of two well substates of different energies being generated from the reflection of the particle standing wave (which itself consists of waves traveling in opposite directions with different energies). That is, the well substate is a superposition in which the particle substate traveling both to the right and left have reflected, delivering different energies to the well.

\section{Discussion}
\label{sec:discussion}

Total destructive correlated interference of the two-body wavegroups, traveling in opposite directions in the $(x_{2},x_{1})$ plane, corresponds to positions where the particle and well or barrier can never be found. One method to verify this effect is to measure the cm of the particle and well or barrier with instruments which have a spatial resolution that is smaller than the fringe spacing along both coordinates. For a static barrier this spacing is half the deBroglie wavelength of the particle, which at $5000$ \AA~for ultracold atoms \cite{cronin} satisfies this requirement while it is dubious at $1.4$ \AA~for slow neutrons \cite{pushin}.

Another constraint on the interference is that the fringe visibility function must be non-zero. That is, the incident and reflected two-body wavegroups must `overlap' within approximately a longitudinal coherence length \cite{coherencelength}, which is given by $l_{c} \approx \lambda^{2}/\Delta \lambda = \lambda V/\Delta V$ \cite{hasselbach}. For particle substates this can be $l_{c}^{particle}=10000$ \AA~for ultracold atoms \cite{cronin} or $l_{c}^{particle}=790$ \AA~for slow neutrons \cite{pushin}.

In this localized region of type I correlated interference, simultaneous measurement is then confined both to a size determined by the two coherence lengths and a temporal duration given by the time during which the wavegroups overlap. The former is small for macroscopic barrier masses while the duration of the interference is essentially determined by the speed of the particle and its coherence length, $\approx l_{c}^{particle}/v$ for a static barrier when $l_{c}^{particle} >> l_{c}^{barrier}$. 

One method to reduce these coherence length limitations is with type II correlation interference, an example of which is peak $b$ in figs. \ref{fig:barrier_negenergy} and \ref{fig:barrier_posenergy}. In this case the interfering two-body wavegroups travel in the same direction in the $(x_{2},x_{1})$ plane after `reflection' from the two barrier edges and are therefore no longer confined to a small region of space. A practical example is in retro-reflection of a neutron from each of the two surfaces of a moving aluminum `slab'. 

This non-local two-body interference is to be contrasted with the local correlated interference just discussed and is similar to that of a pulse of light retro-reflecting from a thin film where destructive interference depends neither on the locations of the detector nor the thin film. The physical difference between neutron-slab correlated destructive interference and this classical example is that in the former case neither the reflected neutron nor slab can be measured simultaneously whereas in the latter case only the pulse of light can not be measured due to destructive interference. 

A more detailed discussion of such non-local interference and of issues associated with possible potential experimental verification of the correlated interference results presented here is similar to those for a particle reflecting from a mirror \cite{kowalski,kowalski2}. Other similarities discussed in these references and relevant to the interaction of a particle with a well or barrier are: measurement of the particle but not the barrier or well, distortion of the wavegroups upon reflection, transfer of coherence between the particle and well or barrier, and the possibility of extending the quantum-classical boundary to a macroscopic mass using correlated interference.

The two body solution for the particle and well or barrier system, $\Psi[x_{1},x_{2},t_{1},t_{2}]$, is given in terms of its subsystem parameters. Operators for the {\em system} therefore involve combining subsystem operators. For example, the center of mass position operator for the system is given by ${\hat X}_{cm}={\hat x}_{1}+{\hat x}_{2}$. This sum must generate a position eigenvalue which agrees with that expected for the cm motion of the classical system, which under the assumptions given is not influenced by any external potentials. To satisfy these constraints, both subsystem positions must be measured simultaneously since an asynchronous measurement leads to different cm positions determined by the different times the particle and well or barrier are measured. Therefore operators for a system observable must act on $\Psi[x_{1},x_{2},t_{1},t_{2}]|_{t_{1}=t_{2}}$. In all cases, including tunneling through the barrier, the momentum and energy eigenvalues of the system are then real even though the subsystem 
momentum of the particle and barrier may be complex or one of the subsystem energies may be negative. 

These results, although far from being comprehensive, indicate a potential direction for further research in understanding quantum correlation.

\begin{acknowledgments}
The authors would like to thank Professors C. Durfee and J. Scales for useful comments.
\end{acknowledgments}


\begin{references}

\bibitem{Gottfried} K. Gottfried, Am. J. Phys. {\bf 68} 143 (2000).
\bibitem{rauch} H. Rauch, Contemp. Phys. {\bf 2} 345 (1986).
\bibitem{gordon} J. P. Gordon, Phys. Rev. A,{\bf 8} 14 (1973).
\bibitem{loudon} S.M. Barnett and R. Loudon, Phil. Trans. R. Soc.A {\bf 368} 927 (2010).
\bibitem{kowalski} F.V. Kowalski and R.S. Browne, arXive, 1404.7160, quant-ph. (2014).
\bibitem{kowalski2} F.V. Kowalski and R.S. Browne, arXive, 1405.0031, quant-ph. (2014).
\bibitem{yin} J. Yin, et al., Phys. Rev. Lett. {\bf 110}, 260407 (2013).
\bibitem{Mcguire} J.H. McGuire and A.L. Godunov, Phys. Rev. A, {\bf 67} 042701 (2003).
\bibitem{currentdensity} In general, the incident and reflected wavevectors for the particle and well or barrier are not equal and opposite as they are for a one particle system in which the barrier or well is fixed. The total probability current density to the left of the barrier then exhibits spatial and temporal interference due to the effect this difference in wavevectors has on the incident and reflected wavefunctions. One consequence is a non-additive current density: e.g. for the barrier potential, the incident and reflected current densities do not simply sum to the transmitted current density. A simple interpretation is that the space to the left of the barrier is increasing as the barrier moves. The net flux  must account for the build-up of probability within this ``new'' space.
\bibitem{Wiener} O. Wiener, Ann. d. Physik {\bf 40}, 203 (1880).
\bibitem{serge} C.U.Serge and J.D. Sullivan, Am. J. Phys. {\bf 44} 729 (1976).
\bibitem{cronin} A. D. Cronin, J. Schmiedmayer, and D. E. Pritchard, Rev. Mod. Phys. {\bf 81}, 1051 (2009).
\bibitem{pushin} P.A. Pushin et. al., Phys. Rev. Lett. {\bf 100}, 250404 (2008).
\bibitem{coherencelength} The coherence length does not change even though the size of the wavepacket does. See A.G. Klein, G.I. Opat, and W.A. Hamilton,  Phys. Rev. Lett., {\bf 50}, 563 (1982).
\bibitem{hasselbach} M. Ferrero and A. van der Merwe, \textit{Fundamental Problems in Quantum Physics}, p-125 (Kluwer Academic Publishing, Dordrecht, 1995).

\end{references}
\end{document}